\documentclass[12pt,aps,prd,preprint,tightenlines,superscriptaddress,
nofootinbib]{revtex4}
\newcommand{\PRE}[1]{{#1}} 

\usepackage{hyperref}      
\usepackage{bm}
\usepackage{epsfig}

\newcommand{\eqref}[1]{Eq.~(\ref{#1})}

\newcommand{\bea}{\begin{eqnarray}}
\newcommand{\eea}{\end{eqnarray}}
\newcommand{\beq}{\begin{equation}}
\newcommand{\eeq}{\end{equation}}
\newcommand{\beqa}{\begin{eqnarray}}
\newcommand{\eeqa}{\end{eqnarray}}
\newcommand{\nn}{\nonumber}

\newcommand{\drawsquare}[2]{\hbox{%
\rule{#2pt}{#1pt}\hskip-#2pt
\rule{#1pt}{#2pt}\hskip-#1pt
\rule[#1pt]{#1pt}{#2pt}}\rule[#1pt]{#2pt}{#2pt}\hskip-#2pt
\rule{#2pt}{#1pt}}

\newcommand{\Yfund}{\drawsquare{6.5}{0.4}}
\newcommand{\Yafund}{\overline{\Yfund}}
\newcommand{\Ythreea}{\raisebox{-3.5pt}{\drawsquare{6.5}{0.4}}\hskip-6.9pt%
        \raisebox{3pt}{\drawsquare{6.5}{0.4}}\hskip-6.9pt
        \raisebox{9.5pt}{\drawsquare{6.5}{0.4}}}

\def\kahler{K\"ahler}

\newcommand{\muv}{M_{{\rm UV}}}

\newcommand{\tr}{{\rm Tr}}

\begin{document}

\title{ \PRE{\vspace*{1.5in}} 
Metastable Rank-Condition Supersymmetry Breaking in 
a Chiral Example
\\
\PRE{\vspace*{0.3in}} }

\author{Yael Shadmi\PRE{\vspace*{.5in}}}
\affiliation{Physics Department, Technion-Israel Institute of
Technology, Haifa 32000, Israel
\PRE{\vspace*{.5in}}
}

\date{July 2011}

\begin{abstract}
\PRE{\vspace*{.3in}} 
We discuss generalizations of Intriligator-Seiberg-Shih (ISS) vacua
to chiral models. We study one example, of an s-confining theory,
in detail.
In the IR, this example reduces to two ISS-like sectors, 
and exhibits a supersymmetry-breaking vacuum with all pseudo-moduli 
stabilized at the origin, and with the R-symmetry unbroken.
The IR theory is interesting from the point of view of R-symmetry breaking.
This theory is an O'Raifeartaigh model with all
charges zero or two,
but the presence of a second R-charged pseudo-modulus
with superpotential couplings to the messengers in principle allows
for R-symmetry breaking.
\end{abstract}


\maketitle

\section{Introduction}
\label{sec:introduction}
Many examples of dynamical supersymmetry breaking theories are known,
but these examples are rather non-generic~\cite{ssreview}. 
If one gives up however on the requirement of a global supersymmetry-breaking
vacuum, which is not essential for model building purposes anyway,
the catalog of dynamical supersymmetry breaking theories may be greatly 
expanded.
The discovery by Intriligator, Seiberg, and Shih (ISS) that metastable, 
supersymmetry-breaking vacua 
exist in simple theories such as SQCD~\cite{iss}, suggests that 
local dynamical supersymmetry breaking  may indeed be a generic phenomenon. 
In this note, we take a first step towards searching for ISS-like minima 
in chiral models.
To the best of our knowledge, no such examples are currently known, 
although there is no reason of principle that chiral theories
could not exhibit local supersymmetry-breaking vacua. 

In order to obtain calculable minima in which supersymmetry is 
dynamically broken via rank-conditions,
one must find chiral theories whose IR descriptions involve some 
tensor (under the global symmetry) fields with dynamically-generated
cubic superpotential couplings.
s-confining theories~\cite{seiberg,sconf} provide a good starting 
point for this search,
since they have smooth moduli-spaces and their IR behavior is well understood.
Of the ten classes of s-confining SU(N) theories~\cite{sconf},  
seven are chiral, in the sense that one cannot give mass to all the fields
of the theory. Many of these have fields in 2-index (or higher) representations
of the global symmetry group in the IR, but these fields 
often have only non-renormalizable
superpotential couplings. In fact, only two models allow for 
rank-condition breaking: one is an SU(5) theory with two copies of 
$(10+\bar5)$ plus two extra 
flavors~\cite{mason},
and the other is an  SU(6) theory with a 3-index anti-symmetric tensor
and 4 flavors. 
Here we will focus on the latter because of its simplicity.
Admittedly, this simplicity is related to the fact that 
the anti-symmetric tensor is a pseudo-real 
representation of SU(6), so that the IR theory is similar to a vector-like
theory. 
In fact, the IR theory  essentially consists
of two sets of ISS SQCD, plus one gauge invariant that couples to these
fields only through non-renormalizable terms.
Still, the theory is chiral because no mass term can be given to
the anti-symmetric tensor field, and this field is an essential 
ingredient
of the dynamics that generates the IR structure.

At the cubic level, the IR superpotential of the model contains two parts.
One is identical to the ISS superpotential, with some fields 
getting supersymmetry-breaking mass splittings at tree-level. 
We will refer to these fields
as messengers, in the spirit of GMSB models~\cite{dnns}.
The second part of the superpotential contains
couplings of the remaining pseudo-moduli to the messengers.
As is well known by now~\cite{gkk, sudano}, this results in a
calculable, rising potential at large field VEVs, so that the pseudo-moduli
are stabilized. 

It is not apriori clear however that the remaining pseudo-moduli
are stabilized at the origin.
The IR theory is an O'Raifeartaigh model, with all charges
being 0 or 2. Such models were argued to preserve R-symmetry~\cite{shih},
and indeed, the mass-squared of the supersymmetry-breaking pseudo-modulus is
positive at the origin.
However, this conclusion applies only if one neglects cubic interactions
that do not involve the supersymmetry breaking modulus\footnote{This loophole
was recently pointed out in~\cite{evans} too.}.
These interactions induce masses-squared for the remaining pseudo-moduli
at one-loop, much like matter-messenger couplings
in GMSB models, which are notorious for generating negative contributions
to the scalar masses-squared at 
one-loop~\cite{dine,pomarol,cheng}\footnote{The 
one-loop contributions of matter-messenger couplings to scalar masses 
were mainly calculated in the limit of small 
supersymmetry breaking~\cite{dine,pomarol}.
With the messenger spectrum of Minimal Gauge Mediation they vanish
to leading order in the supersymmetry breaking as explained in~\cite{gr}.
There are also some examples for which
these contributions were calculated for large supersymmetry 
breaking~\cite{cheng}, with either sign possible.}.
Still, in the present example, the resulting masses-squared are positive,
so that the pseudo-moduli are stabilized at the origin with the R-symmetry
unbroken.

It is worth noting that the IR superpotential we study involves two
uncalculable couplings, and, as a result, the masses-squared of some
of the pseudo-mouli at the origin are  given by a non-trivial function 
of the ratio of these couplings.
Surprisingly however, it turns out that the masses-squared
are positive for any value of this ratio.
This seems to hint at some general argument for the sign of the masses,
and it would be interesting to understand this point.

Finally, a nice feature of the  model is that one does not necessarily
have to introduce a small mass scale by hand in order to obtain
a calculable meta-stable minimum. Instead of adding a mass term in the
UV, it possible to add a higher-dimension term which becomes a linear
term in the IR, with a coefficient that is naturally small.

We describe the theory we consider in Section~\ref{su6}~.
In Section~\ref{linm} we study a mass deformation.
We comment on a non-renormalizable deformation in Section~\ref{linphi}~.
Some details of the calculation of the mass-squared are presented
in the Appendix.

\section{The SU(6) theory}\label{su6}
We consider an SU(6) gauge theory with a single 3-index anti-symmetric
tensor ${\cal A}$, and four flavors (see Table). 
The theory was shown in~\cite{sconf} to s-confine.
The IR theory consists of the gauge invariants listed in the Table,
\begin{equation}\nn
\begin{array}{|c|c|c|c|c|c|c|}
\hline
 &SU(6)&SU(4)_L&SU(4)_R&U(1)_1&U(1)_2&U(1)_R\\
\hline
{\cal A} & \Ythreea &1&1&0&-4&-1\\
Q& \Yfund &\Yfund & 1& 1& 3&1\\
\bar{Q}&\Yafund &1&\Yafund&-1&3&1\\
\hline
\hline
M\sim(Q\bar Q)&&\Yfund&\Yafund&0&6&2\\
\Phi\sim(Q {\cal A}^2 \bar{Q} )&&\Yfund&\Yafund&0&-2&0\\
\hline
\bar{V}\sim({\cal A}Q^3)&&\Yafund&1&3&5&2\\
 V\sim({\cal A}\bar Q^3)&$\ $& 1&\Yfund&-3&5&2\\
\bar B\sim({\cal A}^3Q^3)&&\Yafund&1&3&-3&0\\
 B\sim({\cal A}^3\bar Q^3)&$\ $& 1&\Yfund&-3&-3&0\\
\tilde {\cal A}\sim ({\cal A}^4)& $\ $&1&1&0&-16&4\\
\hline
\end{array}\nonumber
\end{equation}
with the superpotential~\cite{sconf},
\beq\label{w}
W=\frac1{\Lambda^{11}}\, \left[
M_{IA} B_I \bar{B}_A +  
\Phi_{IA} \left(V_I \bar{B}_A + B_I \bar{V}_A
\right) 
\right] +W_{NR}\,.
\eeq 
Here $I,A=1,\ldots,4$, $\Lambda$ is the dynamical scale of the theory,
and
\beq
W_{NR}=\frac1{\Lambda^{11}}\, \left[
V M \bar{V} \tilde{{\cal A}} + M\Phi^3 + \tilde{{\cal A}} M^3 \Phi 
\right]
\eeq 
where we suppressed global symmetry indices.

Since we are interested in vacua near the origin, with all field
VEVs much smaller than $\Lambda$, the non-renormalizable part of the
superpotential, $W_{NR}$, can be neglected. 
In order to give mass to the field $\tilde{{\cal A}}$, which only appears in 
$W_{NR}$, we can either add a singlet $S$ to the theory,
with the superpotential coupling
\beq
\Delta W= \frac1{\muv^2}\,
S\,({\cal A}^4)\,,   
\eeq 
where $\muv$ is the UV cutoff scale,
or simply add the superpotential term,
\beq
\frac1{\muv^5}\, ({\cal A}^4)^2\,.   
\eeq
Either one of these becomes a mass term for $\tilde{{\cal A}}$ in the IR.

Following ISS, we could in principle add the superpotential
\beq\label{deform}
W_0 = {m_0}_{IA}\, Q_I \bar Q_A + 
\frac1\muv {\lambda_\Phi}_{IA}  Q_I {\cal A}^2 \bar Q_A\ ,
\eeq
where $\lambda_\Phi$ is dimensionless.
In the IR theory, this becomes
\beq
W_0 = -\left({\mu_M}\right)_{IA} M_{IA} 
- \left({\mu_\Phi}\right)_{IA} \Phi_{IA} \ , 
\eeq
with $\mu_M\sim\Lambda m_0$ and
$\mu_\Phi\sim \Lambda^2/\muv $.
With matrices $\mu_M$ of rank greater than 1 and/or $\mu_\Phi$ of rank greater
than 2, supersymmetry is broken, since the $F$ term for $M_{IA}$ is
\beq
B_I \bar B_A -\mu_{IA}
\eeq
and $B_I  \bar B_A$ has rank 0 or 1.
Similarly the $F$ term for $\Phi_{IA}$ is 
\beq
B_I \bar V_A + V_I \bar B_A -\mu_{IA}
\eeq
and $B_I \bar V_A + V_I \bar B_A$ has maximal rank 2.

In the following, we will consider perturbations that preserve
the maximal possible global symmetry, and therefore take
${\mu_M}$ and $\mu_\Phi$ to be proportional to the identity.
From the point of view of the IR theory, only one combination of $M$ and 
$\Phi$ appears linearly and triggers supersymmetry breaking.
Using the results of ISS, it is easy to see what happens.
The field that appears linearly couples to some combination(s)
of $B$ and $V$. One then finds an ISS-like model involving one combination
of $M$, $\Phi$, and some combination of $B$-$V$
with a supersymmetry-breaking minimum at the origin,
and with all scalars apart from the Goldstones getting mass either 
at tree-level or at one-loop.
The orthogonal combinations of $M-\Phi$ and $B-V$ couple to these through
the superpotential and are therefore stabilized as well~\cite{sudano}.
Here we will focus on the $M$ perturbation for simplicity. 
From the point of view of the microscopic theory however, adding a
linear 
term in $\Phi$ has some aesthetic advantage,
since, in order to have a calculable minimum we want $\mu_M$ and/or $\mu_\Phi$
much smaller than $\Lambda$. 
This automatically holds for $\mu_\Phi$, since the $\Phi$ tadpole
originates from a non-renormalizable term. 

\section{A linear term in $M$}\label{linm}
Consider first adding just a linear term in $M$, with
${m_0}_{IA}\propto \delta_{IA}$ and ${\lambda_\Phi}_{IA}=0$.
Written in terms of canonically-normalized gauge invariants,
the superpotential of the IR theory is
\beq\label{wir}
W=h \mu M_{IA}\left(-\delta_{IA}  +  B_I \bar{B}_A\right) +  
\lambda \Phi_{IA} \left(V_I \bar{B}_A + B_I \bar{V}_A
\right) \,.
\eeq
Here we used the fact that the \kahler~potential of the IR theory starts 
as\footnote{This form follows from the 
SU(4)$^2\times$U(1)$^2\times$U(1)$_R$~global symmetry, 
plus the exchange symmetry $Q\leftrightarrow \bar Q$,
${\cal A}\leftrightarrow {\cal A}^{dual}$, with 
${{\cal A}^{dual}}^{i_1i_2i_3}\equiv
\epsilon^{i_1\ldots i_6} {\cal A}_{i_4 i_5 i_6}$~and with
the vector  superfield changing sign. This exchange symmetry also
guarantees the equality of the coefficients of the last two terms 
of~\eqref{wir}.},
\beq
K= \frac1{\alpha_M^2} M^\dagger M + \frac1{\alpha_\Phi^2} \Phi^\dagger \Phi
+\frac1{\alpha_B^2} \left(B^\dagger B +\bar{B}^\dagger \bar{B}\right)
+\frac1{\alpha_V^2} \left(V^\dagger V +\bar{V}^\dagger \bar{V}\right) +\cdots \,,
\eeq
where the $\alpha$'s are non-calculable order-one coefficients,
and rescaled the fields $M$, $\Phi$, $B$, $V$, defining
\beq
\mu=\frac1{\alpha_B^2}m_0\Lambda\,,\ \ \ 
   h=\alpha_M\alpha_B^2\,,\ \ \  \lambda=\alpha_\Phi\alpha_B\alpha_V\,. 
\eeq
The model consists of two copies of the ISS fields, $(M, B,\bar B)$~ 
and $(\Phi, V,\bar{V})$~ with the same
charges under all the global symmetries apart from $U(1)_2$.
The first two terms of~\eqref{wir} are precisely the superpotential of 
the ISS SQCD model with three colors and and four flavors,
which has a minimum near the origin with all scalars (apart from the 
Goldstone bosons)
in $(M, B,\bar B)$~getting mass either at tree-level or at one-loop.

What happens to the remaining fields  $(\Phi, V,\bar V)$ at this extremum?
To answer this question,  
let us review the details of the ISS minimum. 
At this minimum, $M=0$, and we can choose $B_4=\bar{B}_4=\mu$ up to global
symmetries. Thus, the $F$ term equation for $M_{44}$ is satisfied.
This is the maximal number of $F$-equations that can be satisfied
in this case. 
Using small Latin indices
for the unbroken $SU(3)_D$,
the $F$ terms of $M_{ii}$ with $i=1,2,3$ are nonzero, with 
$F_{M_{ii}}=-\mu^2$. 
The unbroken symmetry at the minimum is $SU(3)_D\times U(1)^\prime\times U(1)_R$
where $U(1)^\prime$ is a combination of $SU(4)_D$ and $U(1)_1$.

It will be convenient for our purposes to describe the spectrum of 
the ISS fields in analogy with minimal gauge mediation (MGM) 
models~\cite{dnns},
splitting the fields according to their $SU(3)_D$ representations.
We thus define, following ISS,
\beqa\label{split}
B_i&=& b_i\,,\ \ \ \bar B_i= \bar b_i \nn\\
B_4&=&\mu + B_4^+ +B_4^-\,,   
\ \ \ \bar B_4=\mu +  B_4^+ - B_4^-\,,\nn\\   
M_{ia} &=& X_{ia} \\ 
M_{4a}&=&z_a \,, \ \ \ M_{i4}=\bar{z}_i \ .\nn
\eeqa

Expanding around the minimum one has, from the first part of~\eqref{wir},
\beqa
W=   h \bar{b} X b
+h\mu (\bar b z+ \bar z b) 
 + h (\bar{z} b  - \bar{b} z) B_4^- 
+ h\mu M_{44} B_4^+ +\cdots
\eeqa
where we omitted cubic terms involving $M_{44}$, $B_4^\pm$. 
The fields $M_{44}$ and $B_4^+$ get mass at tree level.
The $SU(3)_D$ singlet $X\equiv\tr M$ plays the role of the MGM singlet, with
$F_X= h^2\mu^2$, and splits the masses-squared
of the scalars in the messenger fields $b$, $\bar{b}$,  
while the supersymmetric masses of $b$ and $\bar{b}$ arise from
their couplings to $z$, $\bar{z}$.
All in all, the $b-z$  sector contains two fermions of masses
$\pm h\mu$, two scalars of the same masses (from $z$ and $\bar z$), 
and two scalars
(from $b$, $\bar b$) with masses-squared $h^2\mu^2 \pm F_X=0, 2h^2\mu^2$.
The remaining pseudo-moduli $X$ and $B_4^-$ obtain masses at one-loop, through
their superpotential couplings to the messengers.

Let us turn now to the fields $\Phi$, $V$, $\bar{V}$. 
Splitting these according to their SU(3)$_D$ representations
as in~\eqref{split} we write
\beqa\label{split1}
V_i&=& v_i\,,\ \ \ \bar V_i= \bar v_i \nn\\
V_4&=& V_4^+ +V_4^-\,,   
\ \ \ \bar V_4=  V_4^+ - V_4^-\,,\nn\\   
\Phi_{ia} &=& Y_{ia} \\ 
\Phi_{4a}&=&z^\prime_a \,, \ \ \ \Phi_{i4}=\bar{z}^\prime_i \ .\nn
\eeqa
so that the remaining piece of the superpotential~\eqref{wir} takes the form
\beqa\label{wv}
W=  
\lambda\mu\, (\bar{z}^\prime v  + \bar{v} z^\prime) 
+ \lambda\mu\, \Phi_{44} V_4^+ +
\lambda Y \left(\bar b v +\bar v b \right) +
\lambda V_4^- \left(\bar b z^\prime -\bar{z}^\prime b \right) +
\cdots
\eeqa
where we again neglected irrelevant cubic terms.
At tree-level, just as in the ISS sector, the fields $\Phi_{44}$,  
$V_4^+$ and $v-\bar v$, $z^\prime-\bar{z}^\prime$ get mass $\lambda\mu$ 
with $Y$, $V_4^-$ remaining massless.
These pseudo-moduli couple to the messengers through the 
superpotential~\eqref{wv} and are therefore stabilized~\cite{gkk,sudano}. 
The reason is that, far from the origin  
(but for VEVs still smaller than $\Lambda$, where the theory is calculable)
the potential for these fields can be reliably computed 
from the wave-function renormalizations of the light fields,
as in~\cite{gr}. This always results in a rising potential.

To find the masses near the origin, one must compute the Coleman-Weinberg
potential. We present the result for arbitrary $F_X$ in the Appendix.
For $F_X=h^2\mu^2$ one finds,
\beq\label{mass}
m_Y^2 = \frac13 m_{V_4^-}^2
= \frac{1}{16\pi^2}\, 
\frac{\lambda^2 h^2}{(2-\lambda^2/h^2)(1-\lambda^2/h^2)}\,
\left[\frac{\lambda^2}{h^2}\ln \left(\frac{\lambda^2}{h^2}\right) +
2 \left(1-\frac{\lambda^2}{h^2}\right)\ln2
\right]\, \mu^2\,,
\eeq
which is positive for all values of the ratio $\lambda^2/h^2$,
so that these fields are stabilized at the origin.

With the addition
of the linear term in $M$, the IR theory has an R-symmetry under which
$M$ has charge 2 so that $B$ and $\bar{B}$ can be chosen to have R-charge 
0\footnote{This symmetry happens to coincide with the anomaly-free 
U(1)$_R$ symmetry of the UV theory as given in the~Table, 
but for this discussion all
we care about is the effective R-symmetry of the IR theory.}.
Therefore, the charges of $V$ ($\bar V$) and $\Phi$ must sum to 2,
and we can choose $\Phi$ to have R-charge 2 and  $V$ ($\bar V$) to have
R-charge zero or vice versa.
A priori, the cubic superpotential 
couplings of the pseudo-moduli to
the messengers could have generated negative masses-squared
for $\Phi$ (or $V_4^-$) , leading to $R$-symmetry breaking.
As we saw above, this is not the case.
In fact, the result is positive even if we allow an arbitrary 
$F_X\neq h^2\mu^2$ in~\eqref{wv}.

In addition to the $SU(4)_D\times U(1)_1\times U(1)_R$
symmetry of the UV theory, the superpotential~\eqref{wir}
preserves a global U(1), with, for example, $V$, $\bar V$ having charge 
1 and $\Phi$ having charge $-1$. 
The only terms with $d\leq3$ (in the IR fields) consistent with
the IR symmetry are those already appearing in~\eqref{wir}, but 
the singlet and adjoint pieces in $M$ and $\Phi$
could now appear with different coefficients.
Thus for example, one could add the term $\tr \Phi B \bar V$,
or, (just as in the  ISS model) $M_{adj} B \bar B$, where $M_{adj}$ is
the $SU(4)$ adjoint part of $M$.
The superpotential of the IR model is therefore not generic.
However, if these terms are added by hand in the UV theory,
they are suppressed by the UV cutoff scale, and their contributions
to the masses are therefore smaller than
the radiatively-generated contributions.

 \section{A linear term in $\Phi$}\label{linphi}
Let us briefly comment on adding a linear term in $\Phi$ only, 
setting $m_0=0$ and
$\lambda_\phi\neq0$ in~\eqref{deform}. 
It is convenient to define the combinations
\beq
p_I\propto V_I+B_I\,,\ \ \ q_I\propto V_I-B_I\,,
\eeq
and similarly for the barred fields.
The superpotential then takes the form
\beq
W= \lambda \Phi_{IA}\left( p_I \bar p_A + q_I \bar q_A -\mu_\Phi\delta_{IA}\right)+
\lambda M_{IA}\left( p_I \bar p_A + q_I \bar q_A + 
p_I \bar q_A - q_I \bar p_A  \right)\,,
\eeq
and the supersymmetry breaking scale, $\mu_\Phi\sim \Lambda^2/\muv$ 
is naturally small.
In this case, two of the $F$-term equations for $\Phi$ can be solved,
with two entries of $p$, $q$ getting VEVs. The  pseudo-moduli will again be 
stabilized at one-loop by the superpotential couplings to the messengers.
It would be interesting to study the fate of the $R$ symmetry in
models that contain both the $\tr M$ and the $\tr \Phi$ perturbations.

\section{Conclusions}
As Intriligator, Seiberg, and Shih~\cite{iss} demonstrated, meta-stable 
supersymmetry-breaking vacua appear in the simplest theories.
This suggests that local supersymmetry breaking vacua may occur quite
generically, in both chiral and non-chiral theories. 
In this note, we discussed generalization of the ISS vacua to
chiral theories
in s-confining examples, which are particularly easy to analyze.
It would be interesting to go beyond s-confinement, 
and to explore chiral theories that have weakly coupled IR duals 
in the search for local supersymmetry breaking vacua, whether they are obtained
by rank-condition breaking or by other means. Even chiral theories with 
global, supersymmetry breaking vacua, might posses
additional, local minima with novel features.

\section*{Acknowledgments}
We thank J.~Mason, V.~Kaplunovsky, A.~Katz, Z.~Komargodski, Y.~Shirman and
W.~Skiba for useful discussions. 
We are  grateful to D.~Shih for comments on the draft,
and for bringing reference~\cite{evans} to our attention. 
We are grateful to T.~Lin, J.~Mason, and A.~Sajjad 
for sharing their results with us prior to publication,
and most importantly, for finding a critical sign error 
in our result~\eqref{mass} in the original draft.
We also thank the organizers of 
SUSY Breaking~2011 for a stimulating
workshop. Research supported by the Israel Science
Foundation (ISF) under grant No.~1155/07, and by the United States-Israel
Binational Science Foundation (BSF) under grant No.~2006071.

\appendix
\section{The Coleman-Weinberg calculation}
It is instructive to calculate the 1-loop masses for arbitrary $F$.
We can write the one-loop
correction to the vacuum energy as
\beq
V_{1-loop} = \frac1{64\pi^2}\, \left(
\tr m_B^4 \log \frac{m_B^2}{\Lambda^2} - \tr
\bar m_B^4 \log \frac{\bar m_B^2}{\Lambda^2} 
\right)
\eeq
where $m_B$ denotes the boson masses, and  $\bar m_B$ denotes
the same mass in the supersymmetric limit ($F=0$).
To obtain this result we used the fact that the correction vanishes
for $F=0$, and that the fermion masses are $F$-independent.

To derive the $Y$ mass we write $m_B^2= a + b \vert Y\vert^2$
neglecting higher orders in $Y$. 
The vanishing of the supertrace gives
\beqa
\Sigma_i a_i &=& \Sigma_i \bar a_i\\
\Sigma_i b_i &=& \Sigma_i \bar b_i\\
\Sigma_i a_i b_i &=& \Sigma_i \bar a_i\bar b_i
\eeqa
where again, bars denote the supersymmetric quantities.
The last equality implies that $V_{1-loop}$ is finite.
The $Y$ mass-squared is then
given by
\beq
m_Y^2 =  \frac1{32\pi^2}\Sigma_i \left( a_i b_i \log a_i
-  \bar a_i \bar b_i \log \bar a_i\right)\,.
\eeq
One then finds
\beqa\label{massf}
m_Y^2 &=&
\frac1{16\pi^2}\,
\frac{h^2 \lambda^2 \mu^2}{ \left( 1-r-f \right) 
 \left( 1-r+f \right)  \left( 1-r \right) }
 \Bigg[ -2\,r^2\ln  \left( r \right) f^2 \nn \\
&+& \left( 1-r
 \right)  \left(  ( 1+f)^2 \left( 1-r-f \right) \ln 
 \left( 1+f \right) + \left( 1-f \right)^2 \left( 1-r+f \right) 
\ln ( 1-f)  \right)  \Bigg] 
\eeqa
with $r=\lambda^2/h^2$, $f=F/(h^2\mu^2)$,
which reduces to~\eqref{mass} for $f=1$.
We note that~\eqref{massf} is positive for all values
of $f\leq 1$.

We also note that, unlike the one-loop contributions of matter-messenger
couplings in MGM models, this contribution does not vanish
at $O(F^2)$, because the supersymmetric mass of the messengers 
does not arise from $X$.


\providecommand{\href}[2]{#2}\begingroup\raggedright\endgroup

\end{document}